\documentclass[aoas,nameyear,dvips]{arximspdf}
\usepackage{graphicx}
\usepackage{breakurl}

\doi{10.1214/10-AOAS378}
\volume{5}
\issue{1}
\pubyear{2011}
\firstpage{283}
\lastpage{308}

\makeatletter
\renewcommand{\epsilon}{\varepsilon}

\newproclaim{remark}{Remark}

\newcommand{\binom}{\operatorname{Binomial}}
\newcommand{\numwords}{63}
\newcommand{\T}{\mathcal T}
\newcommand{\bY}{\mathbf Y}
\newcommand{\bx}{\mathbf x}
\newcommand{\betah}{\hat\beta}
\newcommand{\half}{\frac{1}{2}}

\makeatother

\begin{document}
\begin{frontmatter}

\title{Detecting multiple authorship
of United States Supreme Court legal decisions using function words}
\runtitle{Supreme Court decisions}

\begin{aug}
\author[A]{\fnms{Jeffrey S.} \snm{Rosenthal}\corref{}\ead[label=e1]{jeff@math.toronto.edu}}
\and
\author[B]{\fnms{Albert H.} \snm{Yoon}}

\runauthor{J. S. Rosenthal and A. H. Yoon}

\affiliation{University of Toronto}

\address[A]{Department of Statistics and Faculty of Law \\
University of Toronto \\
Toronto\\
Canada\\
\printead{e1}} 
\end{aug}

\received{\smonth{12} \syear{2009}}
\revised{\smonth{5} \syear{2010}}

\begin{abstract}
This paper uses statistical analysis of function words used in legal
judgments written by United States Supreme Court justices, to determine
which justices have the most variable writing style (which may indicated
greater reliance on their law clerks when writing opinions), and also the
extent to which different justices' writing styles are distinguishable
from each other.
\end{abstract}


\end{frontmatter}

\section{Introduction}

      This paper describes various statistical analyses performed on the
texts of judicial opinions written by United States Supreme Court (USSC)
justices.

      Our primary motivation is to attempt to use textual analysis to
explore issues of authorship.  With respect to the USSC, we are interested
in the extent to which justices rely on their clerks when writing
opinions.  According to legal scholar and jurist Richard A. Posner,
``Americans\ldots\   could not care less whether Supreme Court justices
or any other judges write their own opinions or have their clerks write
them, provided the judges decide the outcome'' [Posner (\citeyear{posner}), page 143].
While reasonable minds may disagree with Posner on either positive or
normative grounds, it is clear that
the content of judicial opinions matter, particularly at the USSC.
Lower courts are bound by both the holding and the reasoning of USSC
opinions, and litigants---actual and prospective---act in the shadow of
these opinions [Mnookin and Kornhauser (\citeyear{mnookin})].

      The issue of judicial authorship is important from an institutional
and policy perspective.  At the federal level, the expansion in the number
of cases has significantly outpaced the increase in authorized judgeships.
This trend is most pronounced at the USSC, where over the past fifty
years the court's caseload has steadily
increased while the number of justices---and law clerks---has remained
constant.  Justices are asked to handle more work with the same resources.
Given the public and Congressional scrutiny given to the selection of
USSC nominees [Carter (\citeyear{carter})], it is daunting to consider that much of the
Court's work is done by ``twenty-five-year-old law clerks fresh out of
law school'' [Posner (\citeyear{posner2}), page~567].

Given these increased demands, and heterogeneity in justices'
ability and effort, one would expect that justices differ in how they
manage their workload.  Some justices respond, as many have, by joining
the ``cert pool,'' where justices work collectively to evaluate which
cases to grant \textit{certiorari} [Ward and Weiden~(\citeyear{ward}), page 117].  Some may
delegate more of opinion writing, the primary output from the Court, to
their law clerks.  The degree to which the latter has occurred remains a
topic of intense debate by journalists and scholars alike
[Woodward and Armstrong~(\citeyear{woodward}); Lazarus (\citeyear{lazarus}); Toobin~(\citeyear{toobin})].

This debate has been motivated in part by the fact that justices, as
Article III judges, enjoy lifetime tenure.  Critics of lifetime tenure
contend that justices often serve well beyond their productive years
[Garrow (\citeyear{garrow}); Calbresi and Lindgren~(\citeyear{calabresi})].  If true, one
manifestation would likely be increased delegation of work to law clerks as
justices' ability and desire to do the work diminishes over time.  Our
methodology provides a credible statistical approach to examine the
relationship between justices and their clerks, specifically how justices
vary, both with respect to one another and over the court of their tenure
on the USSC.

Our central intuition is that the greater the delegation in the
opinion-writing process, the more heterogeneous the writing style.  At the
extremes, a justice who wrote his own opinions would presumptively have
a more distinct writing style than another justice who relied heavily on
his law clerks.  This view is supported by recent scholarship analyzing
justices' draft opinions where available, finding that a justice who was
more involved in the opinion-writing process produced more textually
consistent opinions than a justice who delegated more of the work to
his clerks [Wahlbeck et al. (\citeyear{wahlbeck}), page 172].  The institutional design of
USSC clerkships provides an additional exogenous mechanism to test
this hypothesis: USSC clerkships are usually for a single term, from
October through August of the subsequent year. Accordingly, the cohort
for law clerks changes in predictable fashion, allowing an examination
of justices' writing style within and across terms.

Although this paper focuses on statistical methodology, the question
of judicial authorship is an important one in both political science
and law.  USSC opinions reflect a principal--agent relationship between
the justices and their clerks. As with any principal--agent relationship,
the degrees to which the clerks' interests correspond with the justices'
depend on their incentives and degree of oversight.  But to even approach
this question requires first a more complete understanding of judicial
authorship. While it may be possible to tackle this assignment by reading
every Supreme Court opinion, our discussions with current USSC scholars
suggest that differences in writing variability across justices may be too
subtle to discern manually. Our paper provides a more systematic approach.
[In a different direction, scholars have recently taken a textual
analysis approach to the USSC in the context of oral
argument; Hawes, Lin and Resnik (\citeyear{hawes}).]

For example, it is believed that within the current USSC,
certain justices [e.g., Scalia;
see Lazarus (\citeyear{lazarus}), page 271] primarily write their own legal decisions,
while others [e.g., Kennedy; see Lazarus (\citeyear{lazarus}), page 274] rely more on
their law clerks to do much of the writing.  While anecdotes abound on
these claims [Peppers (\citeyear{peppers}); Ward and Weiden (\citeyear{ward})],
there are few hard facts about this
and it is mostly a matter of speculation.

We attempt to verify this hypothesis by measuring the variability of
writing style of the majority opinions written by various justices. We
find, using the Kennedy--Scalia example, that Kennedy opinions have
significantly greater variability than do those by Scalia, measured using
various statistics involving the frequencies of various function words (as
described below). Furthermore, using a bootstrap approach, we confirm that
these differences are statistically significant at the 5\% level.  Given our
assumption that greater reliance on clerks produces greater variability of
writing style, this conclusion would appear to provide compelling evidence
that Kennedy does indeed get more writing assistance from law clerks than
does Scalia. We similarly find that Stevens and Souter have significantly more
variability than Scalia, while Rehnquist and Breyer and Thomas have
significantly less variability than Kennedy.

Our secondary motivation is to attempt to identify authorship, solely
by use of word frequencies. Our informal enquiries with USSC constitutional
scholars indicate that they do not believe they are able to do this.
Nevertheless, in this paper we consider various approaches (naive Bayes
classifier, linear classifier), and show using a cross-validation approach
that such algorithms can indeed predict authorship in pairwise comparisons
with accuracy approaching 90\%.  While this determination is superfluous for
authored opinions, it does provide a clear measure of the extent to which
justices have identifiably distinct writing styles from one another.
Moreover, our approach has other relevant applications, such as identifying
the likely author for per curiam opinions (for which no justice is listed
as the author).

Our methodology thus appears to provide useful methods both
for determining multiple authorship and for identifying the recorded
authorship, solely using function words---at least for USSC decisions
(and perhaps beyond). Our analysis required writing extensive software,
which is freely available [Rosenthal (\citeyear{software})] for purposes
of reproducing or extending our results. Further details are given herein.

\subsection{Background on the United States Supreme Court}

The USSC is the highest court in the United States, providing
the final word on all federal issues of constitutional and statutory
law. It has a predominantly discretionary docket, granting \textit{certiorari}
only for ``cases involving unsettled questions of federal constitutional
or statutory law of general interest'' [Rehnquist (\citeyear{rehnquist}), page 238]. While
the USSC is not unique in issuing opinions or creating precedent, its
position at the apex of the judicial hierarchy ensures that practitioners,
legal scholars, and law students closely scrutinize its
opinions.

Each year the Court receives thousands of petitions for \textit{certiorari}
(requests to hear the case), for which it decides which cases to hear, and
issue opinions.  In the 2008--2009 term, for example, the Court received 8966
cert petitions, heard oral argument for 87 cases, and issued 86 opinions
[Judicial Business of the United States Courts (\citeyear{justia}), Table A1].  These
figures reflect a historical trend in which the caseload demands of the
Court has steadily increased.

The workload is considerable.  Unlike the executive and legislative
branches of the federal government, the USSC is administratively lean. The
court itself consists of only nine justices. The chambers of each justice
typically consist of one secretary and four law clerks.

      Justices routinely serve on the USSC well past typical retirement
age or after they vest in their pension (which usually occurs at age 65),
often leaving only due to death or illness. Perhaps in part due to the
Court's tradition of longevity and hard work, Americans consistently rank
the USSC as the most trusted branch of government [Jamieson and
Hennessy (\citeyear{jamieson})]. Justice
Louis Brandeis once commented, ``[t]he reason the public thinks so much
of the Justices of the Supreme Court is that they are almost the only
people in Washington who do their own work'' [O'Brien~(\citeyear{obrien}), pages
12--13].

      At the same time, the USSC remains one of the least understood
branches. Unlike Congress, the USSC deliberates in private. The
deliberations result in a single public document: the opinion itself.
Accordingly, the process by which the USSC produces each opinion remains
largely unknown.  Prior to the 1950s, justices performed most of the
substantive requirements of the job, including writing opinions
[Peppers (\citeyear{peppers}), page 208], while law clerks performed mostly administrative
tasks. As the job demands increased over time, however, without a
corresponding increase in the number of justices, justices relied more
on law clerks to prepare \textit{certiorari} and oral argument memos, as well
as draft and edit opinions [Peppers~(\citeyear{peppers}), page 151]. While the prestige
of a Supreme Court law clerkship is well accepted within legal circles
[Kozinski (\citeyear{kozinski})], the clerks' contribution to their respective justices
remains largely anecdotal
[Peppers (\citeyear{peppers}); Ward and Weiden (\citeyear{ward}); Lazarus (\citeyear{lazarus}); Woodward and
Armstrong (\citeyear{woodward})].
Some accounts of the USSC law clerks directly challenge Brandeis's claim.
Justice Thurgood Marshall, when told that his view of the right to privacy
conflicted with one of his earlier opinions, allegedly replied, ``That's
not my opinion, that's the opinion of [a~clerk from the prior term]''
[Woodward and Armstrong (\citeyear{woodward}), page 238].
Overall, the anecdotal evidence suggests
that justices vary in their reliance on law clerks in the drafting and
editing of opinions.

\subsection{Statistical analysis via function words}

Stylometry, the statistical analysis of texts, has a long history,
including applications to the famous Shakespeare authorship question [see,
e.g., Seletsky, Huang and Henderson-Frost (\citeyear{seletsky}); Burns~(\citeyear{burns}); Wikipedia (\citeyear{wikipedia}), though much
of the investigation has involved historical and other nonstatistical
methods], the Federalist Papers [Mosteller and Wallace (\citeyear{mosteller})], and Ronald
Reagan's radio addresses [Airoldi et al. (\citeyear{fienberg1}, \citeyear{fienberg2})].

\begin{table}
\tablewidth=359pt
\caption{The \numwords\  function words used in the present
study}\label{tab1}
\tabcolsep=0pt
\begin{tabular*}{359pt}{@{\extracolsep{\fill}}c@{}}
\hline
  \textit{a, all, also, an, and,
  any,
  are,
  as,
  at,
  be,
  been,
  but,
  by,
  can,
  do,
  down,
  even,
  for,
  from,}\\
  \textit{had,
  has,
  have,
  her,
  his,
  if,
  in,
  into,
  is,
  it,
  its,
  may,more,
  must,
  no,
  not,
  now,
  of,
  on,}\\
  \textit{one,
  only,
  or,
  our,
  so,
  some,
  such,
  than,
  that,
  the,
  their,
  then,
  there,
  things,
  this,
  to,}\\
  \textit{up,
  was,
  were,
  what,
  when,
  which,
  who,
  with,
  would}\\
\hline
\end{tabular*}
\end{table}

One challenge with statistical textual analysis is separating those
writing features pertaining to writing style, from those pertaining to
specific subject matter content.  To deal with this, a number of authors
[e.g., Fries (\citeyear{fries}); Miller, Newman and Friedman (\citeyear{miller}); Mosteller and Wallace (\citeyear{mosteller});
Airoldi et al. (\citeyear{fienberg1}); Wyatt (\citeyear{wyatt}); Argamon and Levitan (\citeyear{argamon});
see also the lengthy bibliography in Mosteller and Wallace (\citeyear{mosteller})]
have made extensive use of so-called \textit{function words}, that is, words
such as \textit{all}, \textit{have}, \textit{not} and \textit{than}, whose usage
frequencies are thought to be largely independent of the subject matter
under discussion.  Previous studies [Wyatt (\citeyear{wyatt}); Argamon and Levitan
(\citeyear{argamon})] have found
these function words to be of some use, albeit limited, in determining
authorship of disputed writings.  And, as summarized by Madigan et al.
(\citeyear{madigan}),
``The stylometry literature has long considered function words to be
topic-free in the sense that the relative frequency with which an author
uses, for example, ``with,'' should be the same regardless of whether
the author is describing cooking recipes or the latest news about the
oil futures market.''  In any case, such function words appear to be a
useful starting point for content-independent statistical
analysis.

In particular, in their study of the Federalist Papers,
Mosteller and Wallace~[(\citeyear{mosteller}), page 38,
Table 2.5-2] produce a list of 70 function words, culled for their
purposes from certain earlier studies [Fries (\citeyear{fries}); Miller, Newman and Friedman~(\citeyear{miller})].
This list provides the basis for our statistical
analysis, though to improve stability we eliminated the seven function
words (\textit{every}, \textit{my}, \textit{shall}, \textit{should}, \textit{upon}, \textit{will}, \textit{your}) that occur
with frequency less than 0.001 in the USSC judgments, leaving us with
$\numwords$ function words (Table \ref{tab1}).  (We also considered adding
\textit{while} and \textit{whilst}, which Mosteller and Wallace also found to be
very useful, but they too had frequency less than 0.001 in the USSC judgments.
In any case, our results changed very little upon adding or removing
these few words.)

Of course, it is also possible to consider larger-scale features (e.g.,\
sentence length, paragraph length, multi-word phrases) and smaller-scale
features (e.g., frequency of commas or semi-colons, or the letter `e')
and, indeed, our software [Rosenthal (\citeyear{software})] computes some of these quantities
as well.  However, we have found that these additional quantities did not
greatly improve our predictive power (since their frequencies tend to be
similar for different judgments) and, furthermore, it is subtle (due to
differing scales) how best to combine such quantities with function word
frequencies into a single variability measure, so we do not use them here.
[This decision is partially reinforced by Madigan et al. (\citeyear{madigan}), who tried
13 different feature sets for an authorship identification problem,
and found that function words was tied for second-lowest error rate,
just marginally behind a different ``three-letter suffix'' approach.]
Hence, for simplicity and to allow for ``cleaner'' theory, in the present
study we compute using function words only.

\subsection{Data acquisition}

Our data consisted of the complete text of judgments of the USSC from
1991--2009, as provided by the Cornell Law School web site [Cornell (\citeyear{cornell})].
For consistency, we primarily considered the majority opinions written
by the various USSC justices, though we do briefly consider dissenting
opinions below as well.  (While expansive, the Cornell data source
occasionally introduces transcription errors and furthermore apparently
does not contain quite every USSC opinion, e.g., those by Justice O'Connor
are apparently missing.  We assume, however, that such minor limitations
in the data do not significantly bias our results.)

Although the judgment texts were publicly available [Cornell (\citeyear{cornell})], it was
still necessary to download all the data files from the web, convert them
to simple plain-text format, remove extraneous header and footer text,
sort the judgments by authoring justice and by court session, and index
all the judgments by date written.  The number of judgments to consider,
well over 1000, required writing extensive software [Rosenthal (\citeyear{software})],
consisting of various C programs together with Unix shell-scripts,
to quickly and automatically perform this task.

Using this software, we downloaded and processed and sorted all of the
majority opinion judgments (and also separately the dissenting opinions)
of various USSC justices.  To avoid trivialities, judgments containing
fewer than 250 words were systematically excluded.  The resulting files
were then used as data for all of our statistical work below.

\section{Statistical analyses of word counts}

We suppose that our function words are numbered from $j=1$ to $j=\numwords$.
We further suppose that a given justice has written judgments numbered
from $i=1$ to $i=K$.  Let $w_i$ be the total number of words in judgment
$i$, and let $c_{ij}$ be the number of times that function word $j$
appears in judgment $i$.

\subsection{Word fractions}

Since judgments can differ considerably in their length, the raw
counts $c_{ij}$ by themselves are not particularly meaningful.
One approach is to instead consider the quantities
\[
f_{ij}  =  c_{ij}  /  w_i,
\]
representing the fraction of words in judgment $i$
which are function word $j$.  If the sample standard deviation
$\mathit{sd}(f_{1j},f_{2j},\ldots,f_{Kj})$ is much larger for one justice than
for another, this suggests that the former has a much more variable
writing style.

Unfortunately, this analysis is not entirely independent of such factors
as the length of judgments, the different justices' different
propensities to use different words, etc.
For example, suppose for a given function word $j$,
a given justice has some fixed unknown propensity $p_j$ for using
the function word $j$, independently for each word of each of their
judgments.  In this case, the distribution of the
count $c_{ij}$ of reference word $j$ in judgment $i$
is $\binom(w_i,p_j)$, so that $f_{ij}$ has mean
$p_j$ and variance $p_j(1-p_j)/w_i$, which
depend on the individual propensities $p_j$ and judgment lengths $w_i$.
So, while we could consider calculating the sum of sample standard deviations
\[
V_1  =  \sum_{j=1}^{\numwords} \mathit{sd}(f_{1j},f_{2j},\ldots,f_{Kj}),
\]
and use that as a measure of the variability of writing style across
judgments of a given justice,
such a comparison would not be entirely satisfactory since it
would be influenced by such factors as
$p_j$ and $w_i$, so it would, for example, tend to unfairly assign smaller
variability to justices who tend to write shorter decisions.

One approach to overcoming this obstacle is to adjust the fractions
$f_{ij}$ to make them less sensitive to $p_j$ and $w_i$.  Specifically,
$f_{ij}-p_j$ has mean 0 and variance $p_j(1-p_j)/w_i$, so this is also
approximately true for $f_{ij}-\mu_j$, where
\[
\mu_j  =  \frac{c_{1j}+c_{2j}+\cdots+c_{Kj}}{w_1+w_2+\cdots+w_K}
\]
is our best estimate of $p_j$.  Hence, approximately, the quantity
\[
r_{ij}  =  w_i^{1/2} (f_{ij}-\mu_j)
\]
has mean 0 and variance $p_j(1-p_j)$, which is independent of
the judgment length $w_i$, suggesting the refined variability estimator
\[
V_2  =  \sum_{j=1}^{\numwords} \mathit{sd}(r_{1j},r_{2j},\ldots,r_{Kj}).
\]
Since $p_j(1-p_j)$ still depends on the
unknown propensity $p_j$, we could further refine the variability
estimator to
\[
V_3  =  \sum_{j=1}^{\numwords} \mathit{sd}(q_{1j},q_{2j},\ldots,q_{Kj}),
\]
where now
\[
q_{ij}  = \frac{w_i^{1/2} (f_{ij}-\mu_j)}
{(\mu_j(1-\mu_j))^{1/2}}
\]
have mean 0 and variance approximately 1, independent of both $w_i$ and
$p_j$.

However, even these refinements are
not entirely satisfactory, since the $\mu_j$ are
not perfect estimates of the propensities $p_j$, and it is difficult
to accurately take into account the additional variability from the
uncertainty in the $\mu_j$.  In particular, if $p_j$ is quite close
to zero (as it often will be), then dividing by $\mu_j$ might be rather
unstable, leading to unreliable results.  (In the most extreme case, if
$\mu_j=0$, then dividing by $\mu_j$ is undefined; we fix this by simply
omitting all terms with $\mu_j=0$ from the sum, but this illustrates
the instability inherent in $V_3$.)  So, we now consider an alternative
approach.

\subsection{A chi-squared approach}

Since in our case the counts $c_{ij}$ are exact, while estimates such as
$\mu_j$ are inexact, this suggests that we instead use the chi-squared
statistic.  Specifically, we consider the value
\[
\mathit{chisq}  =  \sum_{i=1}^K \sum_{j=0}^{\numwords}
\frac{(c_{ij} - e_{ij})^2}{e_{ij}}
 ,
\]
where again $w_i$ is the total number of words in judgment
$i$, and $c_{ij}$ is the number of times that function word $j$
appears in judgment $i$, and now
$c_{i0} = w_i - c_{i1} - \cdots - c_{iK}$
is the number of words in judgment $i$ which are \textit{not} function
words, and
\[
e_{ij}  = w_i \biggl( \frac{c_{1j}+c_{2j}+\cdots+c_{Kj}}{w_1+w_2+\cdots+w_K }\biggr)
\]
is the expected number of times that function word $j$ would
have appeared in judgment $i$ under the null hypothesis that
the total number $c_{1j}+c_{2j}+\cdots+c_{Kj}$ of appearances of
reference word $j$ were each equally likely to occur in any of the total
number $w_1+w_2+\cdots+w_K$ of words in all of
the justice's $K$ judgments combined.

Under the null hypothesis, the $\mathit{chisq}$ statistic
should follow a chi-squared distribution with
$(\numwords+1-1)(K-1) = \numwords(K-1)$
degrees of freedom, hence with mean $\numwords(K-1)$.  (The
``$+1$'' arises because of the $c_{i0}$ terms.)
So, dividing this
statistic by its null mean, we obtain the new statistic
\[
V_4  =  \mathit{chisq}/\mathit{df}  =  \mathit{chisq}  /  \numwords(K-1).
\]
The value of $\mathit{chisq}/\mathit{df}$ should be approximately 1 under the null
hypothesis, and larger than 1 for writing collections which exhibit
greater writing style variability.  In particular, the extent to which
$\mathit{chisq}/\mathit{df}$ is larger than 1 appears to be a fairly reliable and robust way
to estimate writing style variability.

In our experiments below, we report values of each of $V_1$, $V_2$,
$V_3$ and $V_4$, but we concentrate primarily on $V_4$ since we feel it
is the most stable and reliable measure and eliminates many of the
shortcomings of the other three quantities.

\subsection{Variability results}

We developed software [Rosenthal (\citeyear{software})] to compute each of the above
variability statistics $V_1$, $V_2$, $V_3$ and $V_4$ (among other
statistics).  We then applied our software to a variety of USSC
justices' judgments.  The results are in Table \ref{tab2}.  (Unless otherwise
specified, the results are for \textit{majority} judgments written by
that justice.  Also, ``words/judgment'' means the average number of words per
judgment considered.)

\begin{table}
\tabcolsep=0pt
\caption{Variability statistics for selected USSC justices}\label{tab2}
\begin{tabular*}{\textwidth}{@{\extracolsep{\fill}}lcccccc@{}}
\hline
&  \textbf{\# judgments} & \textbf{Words/Judgment} & $\bolds{V_1}$ & $\bolds{V_2}$ &    $\bolds{V_3}$ &
$\bolds{V_4}$\\
\hline
Kennedy       & 147 &5331       &    0.118 &    8.27 &   2709.1 & 4.12
\\
Scalia        & 156 &4536       &    0.113 &    7.18 &   2467.0 & 3.13
\\ [3pt]
Stevens & 148 &5996 & 0.111 & 7.96 & 2856.2 & 3.94 \\
Souter        & 143  &5638 &    0.111 &    7.80 &   2531.0 & 3.77 \\
Ginsburg & 130 &4712 & 0.114 & 7.49 & 2926.4 & 3.66 \\
Thomas & 140 &3877 & 0.128 & 7.64 & 2669.4 & 3.55 \\
Breyer        & 121  &3804 &    0.119 &    7.06 &   2538.7 & 3.31 \\
Rehnquist & 127 &3743 & 0.124 & 7.22 & 2398.5 & 3.22 \\ [3pt]
Stevens dissent & 205 &3202 & 0.147 & 6.75 & 2251.1 & 2.63 \\
Kennedy dissent & \phantom{0}42  &3546 &    0.148 &    6.49 &   2229.8 & 2.51 \\
Scalia dissent & 108 &3410 & 0.141 & 6.64 & 2083.5 & 2.46 \\
\hline
\end{tabular*}
\end{table}

Looking at these results, we
see that Kennedy does indeed have higher writing-style variability than
does Scalia, by each of the four measures, thus apparently confirming
our original hypothesis (see also the next section re statistical
significance).  The other justices mostly fall in between these extremes,
though Souter and Stevens also have very high variability, while Breyer
and Rehnquist have lower variability.

As for the dissenting judgments, we might expect them to have much
smaller writing variability since they tend to be more focused and
also more likely to be written by the justice alone.  This is indeed
confirmed by the measures $V_2$, $V_3$ and $V_4$, but not by $V_1$
which gets tripped up by the fact that dissents tend to be shorter and
$V_1$ does not correct for this.  So, this provides confirmation that
dissent judgments tend to have more consistent writing style, and also
further illustrates why $V_1$ is not an appropriate measure of variability.
(By contrast, Thomas has greater variability than Scalia by all measures,
even though his average judgment length is much shorter.)

\begin{remark*}
Of course, the different $V_i$ are each on a different scale, so it is
meaningless to, for example, compare the value of $V_1$ directly with the value
of~$V_2$.  It is only meaningful to compare the same variability
statistic (e.g., $V_4$) when computed for different collections of
judgments.
\end{remark*}

\begin{remark*}
In all cases the value of $V_4$ is much larger
than it would be under the null hypothesis that the function words are
truly distributed uniformly and randomly.  For example, for Scalia, the
$\mathit{chisq}$ statistic is equal to $3.13 \times \numwords \times (156-1) =
30{,}564.45$; under the null hypothesis this would have the chi-squared
distribution with $\numwords(156-1)=9765$ degrees of freedom, for which the
value 30{,}564.45 corresponds to a $p$-value of about $\exp(-4834.5)$ which is
completely negligible.  So, the null hypothesis is definitively
rejected.  However, we still feel that $\mathit{chisq}$ (or, in particular, the
related quantity $V_4$) is the most appropriate measure of writing-style
variability in this case, even though it no longer corresponds
to an actual chi-squared distribution.
\end{remark*}

\begin{remark*}
Of course,
while a larger $V_i$ value indicates that one justice has a more
variable writing style than another, it does not directly determine
whether the justice relies more heavily on law clerks.  Alternative
explanations include that the justice edits his/her clerks work more
carefully, or that some clerks are better than others at copying their
justice's writing style, or that some justices naturally have a more
variable writing style even when writing entirely on their own.
So, we view the $V_i$ measurements as \textit{one} window into the reliance
of justices on their clerks, but not a completely definitive one.
This issue is considered further in the next section.
\end{remark*}

\begin{figure}[t]

\includegraphics{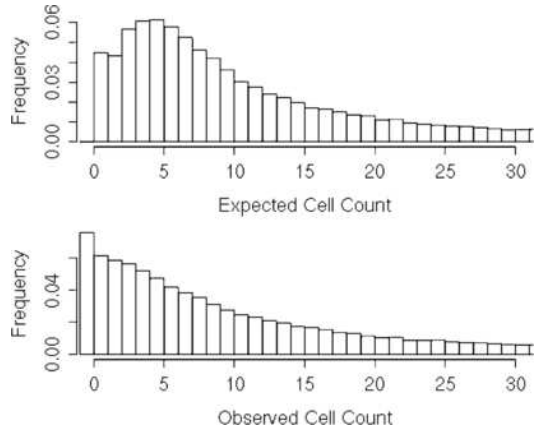}

\caption{Expected (top) and observed (bottom) cell count
frequencies for all 70,056 cells corresponding to
all 1112 of the USSC 1991--2009 judgments considered above.}\label{fig1}
\end{figure}

\begin{remark*}
Values of $\mathit{chisq}$ statistics can be less stable/useful when many of the
expected cell counts are very close to zero.  Despite having already
eliminated from consideration those function words which have very low
frequency in the USSC judgments, and those USSC judgments which are
extremely short, it is still true that in the judgments we consider,
4.48\% of the expected cell counts (mean 28.91, median 9.96) and 7.55\%
of the observed cell counts (mean 28.91, median 10.00) are less than one
(Figure \ref{fig1}).  It may be possible to correct for this, for example, with Yates'
correction, but this is not without difficulties, as it may over-correct
and also is usually used only for $2 \times 2$
tables.  As a check, we experimented
with recomputing our $V_4$ statistic omitting all cells with very small expected
count, and found that this slightly reduced all the $V_4$ values but in
a consistent way, and relevant comparisons such as bootstrap tests of
significance (see next section) were virtually unchanged.  So, overall
we do not expect this small-cell issue to be a significant problem,
and we leave the definition of $V_4$ as above.
\end{remark*}

\section{Bootstrap test of significance}

The question remains whether the results from the previous section
(e.g., that Kennedy has larger writing variability than Scalia) are the
result of mere chance or are actually illustrative of different amounts
of writing-style variability.  Since we have already rejected the null
hypothesis (the null hypothesis that the function words are truly
distributed uniformly and randomly), the quantity $V_4$ no longer
follows a chi-squared distribution, so no simple analytic test
of statistical significance is available.

So, to test significance, we perform a \textit{bootstrap}
test.  Specifically, for each justice we shall select cases
$a_1,a_2,\ldots,a_{100}$ uniformly at random (with repetition allowed,
though this could also be done without repetition).  For each such choice
of 100 cases, we shall compute the variability measure $V_4$ as above.
We shall repeat this 1000 times for each justice, thus giving a list of
1000 different possible values of~$V_4$, depending on which list of 100
cases was randomly selected.

If we do this for two different justices, say, for Kennedy and for Scalia,
then this gives us $1000 \times 1000 = 1{,}000{,}000$ pairs of $V_4$ values.  We then
simply count the fraction of pairs under which the $V_4$ for Kennedy is
larger than the $V_4$ for Scalia, to give us an estimate of the {\it
probability} that $V_4$ for Kennedy is larger than $V_4$ for Scalia,
for a randomly-chosen selection of their judgments.  We also use the
pairs to estimate the distribution function for the difference of the
$V_4$ variability for Kennedy, minus that for Scalia, and then use this
estimated distribution function to compute the 95\% confidence interval
for the difference of $V_4$ for Kennedy minus that for Scalia.  If this
confidence interval is entirely positive, this indicates that Kennedy
judgments have a more variable writing style than Scalia judgments,
and that this conclusion is robust and statistically significant, rather
than merely the result of chance variation.

Note that this bootstrap setup has the additional advantage that,
since the same number of judgments (100) are chosen for each justice at
each step, any concerns about comparing different numbers of judgments
are avoided.

\subsection{Variability bootstrap results}

We developed software [Rosenthal (\citeyear{software})] to compare the $V_4$ bootstrap
values as above, using 1000 bootstrap samples each of size 100.  We then
ran this software to compare Kennedy and Scalia in this manner, obtaining
the following results:\vspace*{5pt}
\begin{center}
{\fontsize{9}{10}\selectfont{
\begin{tabular*}{\textwidth}{@{\extracolsep{\fill}}cccc@{}}
\hline
$\bolds{\operatorname{mean}}$\textbf{(Kennedy)} & $\bolds{\operatorname{mean}}$\textbf{(Scalia)} & \textbf{P(Kennedy $\bolds{>}$ Scalia)}
& \textbf{95\% CI for
Kennedy $\bolds{-}$ Scalia}\\
\hline
4.06 & 3.08 & 0.99996 & (0.48, 1.51) \\
\hline
\end{tabular*}
}}
\end{center}
\vspace*{5pt}
That is, this bootstrap test determines that the probability that a
randomly-selected sample of Kennedy's writings is more variable than a
randomly-selected sample of Scalia's writings is over 99.99\%,
a near certainty.  Furthermore, the 95\% confidence interval (0.48,
1.51) for the difference in variabilities is entirely positive.  So,
we can conclude with confidence that, based on the $V_4$ chi-squared test,
Kennedy's writings have more variable writing style than Scalia's.

Similarly, when comparing Souter to Scalia, we obtain the
following:\vspace*{5pt}
\begin{center}
{\fontsize{9}{10}\selectfont{
\begin{tabular*}{\textwidth}{@{\extracolsep{\fill}}cccc@{}}
\hline
$\bolds{\operatorname{mean}}$\textbf{(Souter)} & $\bolds{\operatorname{mean}}$\textbf{(Scalia)}
& \textbf{P(Souter} $\bolds{>}$ \textbf{Scalia)} & \textbf{95\% CI for
Souter $\bolds{-}$ Scalia}\\
\hline
3.72 & 3.07 & 0.995 & (0.15, 1.20) \\
\hline
\end{tabular*}
}}
\end{center}
\vspace*{6pt}Or, comparing Kennedy's majority opinions to Kennedy's dissents, we obtain the following:\vspace*{5pt}
\begin{center}
{\fontsize{9}{10}\selectfont{
\tabcolsep=3pt
\begin{tabular*}{\textwidth}{@{\extracolsep{\fill}}cccc@{}}
\hline
$\bolds{\operatorname{mean}}$\textbf{(majority)} & $\bolds{\operatorname{mean}}$\textbf{(dissent)}
& \textbf{P(majority $\bolds{>}$ dissent)}
& \textbf{95\% CI for majority $\bolds{-}$ dissent}\\
\hline
4.06 & 2.44 & 1.00 & (1.14, 2.13) \\
\hline
\end{tabular*}
}}
\end{center}
\vspace*{5pt}
Or, comparing Scalia's majority opinions to Scalia's dissents, we obtain the following:\vspace*{5pt}
\begin{center}
{\fontsize{9}{10}\selectfont{
\tabcolsep=3pt
\begin{tabular*}{\textwidth}{@{\extracolsep{\fill}}cccc@{}}
\hline
$\bolds{\operatorname{mean}}$\textbf{(majority)} & $\bolds{\operatorname{mean}}$\textbf{(dissent)}
& \textbf{P(majority $\bolds{>}$ dissent)}
& \textbf{95\% CI for majority $\bolds{-}$ dissent}\\
\hline
3.08 & 2.43 & 0.9997 & (0.29, 1.01) \\
\hline
\end{tabular*}
}}
\end{center}
\vspace*{5pt}
Thus, we conclude with confidence that, as expected, both Kennedy's and
Souter's majority opinion writing are more variable than that of Scalia,
and, furthermore, Kennedy's majority opinion writing is more variable than
his dissent opinion writing (and similarly for Scalia).

Similarly, we can compare other justices to Scalia, as follows:\vspace*{5pt}
\begin{center}
{\fontsize{9}{10}\selectfont{
\begin{tabular*}{\textwidth}{@{\extracolsep{\fill}}cccc@{}}
\hline
\multicolumn{1}{@{}l}{\textbf{mean(Stevens)}} & \textbf{mean(Scalia)} & \textbf{P(Stevens $\bolds{>}$ Scalia)} & \multicolumn{1}{r@{}}{\textbf{95\% CI for
Stevens $\bolds{-}$ Scalia}}\\
\hline
3.86 & 3.08 & 0.998 & (0.25, 1.34) \\
\hline\\ [6pt]
\hline
\textbf{mean(Ginsburg)} & \textbf{mean(Scalia)} & \textbf{P(Ginsburg $\bolds{>}$ Scalia)} & \textbf{95\% CI for
Ginsburg $\bolds{-}$ Scalia}\\
\hline
3.59 & 3.07 & 0.988 & (0.07, 0.99) \\
\hline\\ [6pt]
\hline
\multicolumn{1}{@{}l}{\textbf{mean(Thomas)}} & \textbf{mean(Scalia)} & \textbf{P(Thomas $\bolds{>}$ Scalia)}
& \multicolumn{1}{r@{}}{\textbf{95\% CI for Thomas $\bolds{-}$ Scalia}}\\
\hline
3.48 & 3.08 & 0.972 & ($-$0.01, 0.82) \\
\hline
\end{tabular*}
}}
\end{center}
\vspace*{5pt}
Thus, we can conclude that in addition to Kennedy and Souter,
each of Stevens and Ginsburg
also has greater writing variability
than does Scalia, while
Thomas \textit{may} have greater writing variability
than does Scalia but that assertion is not completely established.
[The conclusion about Stevens may be surprising, since Stevens also has a
reputation for doing his own writing; see Domnarski (\citeyear{domnarski}), page 31.
So, this result may indicate that Stevens actually relied on clerks more
than is generally believed, though of course this evidence is not
completely definitive.]

In the other direction, we conclude that in addition to Scalia, also
Rehnquist, Breyer and Thomas each have statistically significantly less
variability than Kennedy, while\vadjust{\goodbreak} Ginsburg does not:\vspace*{5pt}

\begin{center}
\tabcolsep=0pt
{\fontsize{9}{10}\selectfont{
\begin{tabular*}{\textwidth}{@{\extracolsep{\fill}}cccc@{}}
\hline
&&&\textbf{95\% CI for}\\
\textbf{mean(Kennedy)} & \textbf{mean(Rehnquist)} & \textbf{P(Kennedy $\bolds{>}$ Rehnquist)}
& \textbf{Kennedy $\bolds{-}$ Rehnquist}\\
\hline
4.06 & 3.17 & 0.9998 & (0.37, 1.43)\\
\hline\\ [6pt]
\hline
&&&{\textbf{95\% CI for}}\\
\textbf{mean(Kennedy)} & \textbf{mean(Breyer)} & \textbf{P(Kennedy $\bolds{>}$ Breyer)}
&{\textbf{Kennedy $\bolds{-}$ Breyer}}\\
\hline
4.05 & 3.25 & 0.995 & (0.19, 1.42)\\
\hline\\ [6pt]
\hline
&&&{\textbf{95\% CI for}}\\
\textbf{mean(Kennedy)} & \textbf{mean(Thomas)} & \textbf{P(Kennedy $\bolds{>}$ Thomas)}
& {\textbf{Kennedy $\bolds{-}$ Thomas}}\\
\hline
4.06 & 3.48 & 0.981 & (0.03, 1.15)\\
\hline \\ [6pt]
\hline
&&&{\textbf{95\% CI for}}\\
\textbf{mean(Kennedy)} & \textbf{mean(Ginsburg)} & \textbf{P(Kennedy $\bolds{>}$ Ginsburg)}
&{\textbf{Kennedy $\bolds{-}$ Ginsburg}}\\
\hline
4.06 & 3.57 & 0.948 & ($-$0.10, 1.06) \\
\hline
\end{tabular*}
}}
\end{center}
\vfill\eject
\subsection{Within-justice comparisons}

It is possible to use this same $V_4$ bootstrap approach to compare
different collections of judgments by the same justice.

For example, as justices age, their writings might get less variable
(since they develop a more consistent style) or more variable (if they
come to rely more on their law clerks).  To test this, we perform $V_4$
bootstrap tests, as above, except now comparing a justice's majority
opinions from the 1990s decade, to the same justice's opinions from the
2000s decade.  Our results are in Table \ref{tab3}.

\begin{table}[t]
\caption{Bootstrap results comparing the decades 1990s and 2000s}\label{tab3}
\begin{tabular}{@{}lcccc@{}}
\hline
\textbf{Justice} & \textbf{mean(1990s)} & \textbf{mean(2000s)} & \textbf{P(1990s $\bolds{<}$ 2000s)} & \textbf{95\% CI for
2000s $\bolds{-}$ 1990s}\\
\hline
Kennedy & 3.78 & 4.15 & 0.875 & ($-$0.27, 0.96) \\
Scalia & 2.96 & 3.08 & 0.753 & ($-$0.25, 0.47) \\
Souter & 3.73 & 3.49 & 0.189 & ($-$0.79, 0.27) \\
Stevens & 3.78 & 3.82 & 0.539 & ($-$0.57, 0.66) \\
Breyer & 3.37 & 3.05 & 0.138 & ($-$0.86, 0.28) \\
Ginsburg & 3.30 & 3.69 & 0.948 & ($-$0.08, 0.86) \\
Thomas & 3.36 & 3.47 & 0.701 & ($-$0.29, 0.49) \\
Rehnquist & 3.20 & 2.92 & 0.050 & ($-$0.64, 0.06) \\
\hline
\end{tabular}
\end{table}

Looking at these results, there is no clear pattern.  None of the decade
differences are statistically significant.  Rehnquist is \textit{nearly}
significantly more variable in the 1990s than the 2000s, and Ginsburg is
\textit{nearly} significantly more variable in the 2000s than the 1990s, but
since this does not conform to any obvious interpretation or ``story,''
we are inclined to regard these slight differences as mere chance events.

Another way to compare a justice's writing is to look at those judgments
which were in the first half of a session (i.e., September through March)
versus those judgments in the second half (i.e., April through August).
The reason why judgments early in a session may appear different from
those later in a session is because law clerks rotate annually; thus,
writing variability over the course of a session may increase if a given
justice delegates more work to his clerks, or may diminish if clerks
better learn the preferences of their justice.  That is, increasing
variability may indicate a justice's increased trust, and therefore
increased delegation or lower oversight to the clerk; conversely,
decreasing variability could reflect increased understanding by the
clerks of their justice's preferred writing style.

Our results for this comparison are in Table \ref{tab4}.

\begin{table}[t]
\caption{Bootstrap results comparing the first and second halves of court
sessions}\label{tab4}
\begin{tabular*}{\textwidth}{@{\extracolsep{\fill}}lcccc@{}}
\hline
\textbf{Justice} & \textbf{mean(first)} & \textbf{mean(second)} & \textbf{P(first $\bolds{<}$ second)}
& \textbf{95\% CI for
second $\bolds{-}$ first}\\
\hline
Kennedy & 3.68 & 4.25 & 0.964 & ($-$0.05, 1.21) \\
Scalia & 3.04 & 3.05 & 0.521 & ($-$0.40, 0.41) \\
Souter & 3.39 & 3.91 & 0.955 & ($-$0.08, 1.08) \\
Stevens & 4.04 & 3.61 & 0.094 & ($-$1.07, 0.20) \\
Breyer & 3.15 & 3.21 & 0.602 & ($-$0.49, 0.59) \\
Ginsburg & 3.57 & 3.47 & 0.340 & ($-$0.55, 0.35) \\
Thomas & 3.42 & 3.44 & 0.530 & ($-$0.38, 0.41) \\
Rehnquist & 3.01 & 3.28 & 0.935 & ($-$0.08, 0.62) \\
\hline
\end{tabular*}
\end{table}

This time, it appears that Kennedy, Souter and
Rehnquist are somewhat more variable in the \textit{second}
half of court sessions, which is consistent with the hypothesis
that they let their clerks write more opinions once they have
more work experience.  Meanwhile, Stevens leans slightly in the opposite
direction, with more variability in the \textit{first} half of court
sessions.  However, none of these results are statistically
significant, so we refrain from drawing clear conclusions from them.

\section{Further investigation of the ``$V_4$'' statistic}

Since the ``$V_4$'' quantity is central to our conclusions about text
variability and multiple authorship, it seems appropriate to better
understand the performance of this quantity in other circumstances, as
we do now.

\subsection{Randomly-generated text}

As a simple first experiment, we randomly generated 200 pseudo-documents
each consisting of 2000 independent randomly generated words.
(Specifically, each word was chosen to be a nonfunction word with
probability 70\%, or uniformly selected from the list of function
words with probability 30\%.)  For such randomly-generated text, the
$V_4$ statistic should approximately equal 1.  In fact, we repeated this
experiment 10 times, obtaining a~mean $V_4$ value of 1.004622, with a
standard deviation of 0.01701969, consistent with having a true mean
value of 1.

\subsection{Historical trend}

It is generally believed that USSC justices rely more on their clerks
in the modern era than they did in earlier times [Ward and Weiden~(\citeyear{ward});
Peppers (\citeyear{peppers})].
If so, and if larger $V_4$ values are indeed a good indicator of increased
authorship, then $V_4$ values should be generally increasing with time.

\begin{figure}[t]

\includegraphics{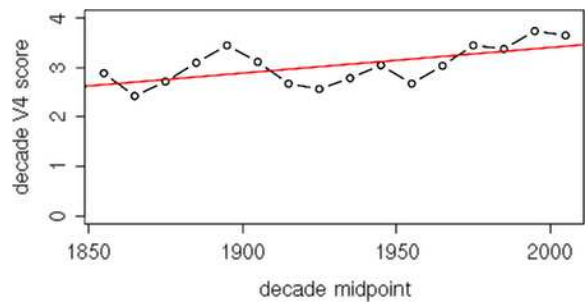}

\caption{Decade-by-decade USSC judgment average $V_4$ scores.}\label{fig2}
\end{figure}

To test this, we extended our software [Rosenthal (\citeyear{software})] to also download
USSC cases from the Justia.com web site [Justia (\citeyear{justia})], and used this to
analyze cases from previous eras.  We then computed the $V_4$ score for all
cases, by all justices, on a decade-by-decade basis from 1850 onward (i.e.,
for all cases decided in 1850--1859, and for all cases decided in
1860--1869 and so on).  The results, plotted against decade midpoint
together with a line of best fit, were as follows (Figure \ref{fig2}).

This graph shows that the $V_4$ scores have
a general trend upward, increasing by just over 0.005 per decade (and this
increase is statistically significant, $p=0.0105$).  This upward trend
is accelerated in the modern era (1950--2009) to 0.020 per decade
($p=0.0087$), corresponding to the period of increasing clerk
activity [Peppers (\citeyear{peppers})].  These observations are
consistent with the supposition that $V_4$ scores increase with increased
delegation of authorship, and that this delegation (through the
greater reliance on law clerks) has increased over the years.

\subsection{Combining justices}

Another way to test whether the $V_4$ statistic increases with multiple
authorship is to combine various collections of judgments together in
ways which would appear to increase the total number of authors, and see
if the $V_4$ scores rise.  We select judgments that
we believe to be homogeneous in their authorship, namely,
majority judgments of Scalia, Rehnquist and Breyer, and dissenting
judgments of Scalia and Kennedy.  Our results are in Table \ref{tab5}.

\begin{table}[t]
\tablewidth=245pt
\caption{Variability statistics when combining different justices together}\label{tab5}
\begin{tabular*}{245pt}{@{\extracolsep{\fill}}lcc@{}}
\hline
&  \textbf{\# judgments} & $\bolds{V_4}$ \\
\hline
Scalia        & 156 & 3.13 \\
Rehnquist & 127 & 3.22 \\
Breyer        & 121 & 3.31 \\ [3pt]
Scalia $+$ Rehnquist & 283 & 3.29 \\
Scalia $+$ Breyer & 277 & 3.44 \\
Rehnquist $+$ Breyer & 248 & 3.49 \\
Scalia $+$ Rehnquist $+$ Breyer & 404 & 3.46 \\ [3pt]
Scalia dissent & 108 & 2.46 \\
Kennedy dissent & \phantom{0}42  & 2.51 \\ [3pt]
Scalia dissent $+$ Kennedy dissent & 150 & 2.56 \\
\hline
\end{tabular*}
\end{table}

This table shows that, while the effect is not overwhelming, nevertheless,
there is a modest increase in the values of $V_4$ when two or three
different justices (each believed to author their own judgments) are
combined together, as compared to the $V_4$ scores for individual
justices.  (Furthermore, this effect appears to be reasonably
robust to subselection.  For example, we divided the Rehnquist and
Breyer opinions into two subcollections and measured the $V_4$ score for the
four possible pairings, obtaining values of 3.37, 3.43, 3.52 and 3.60,
all significantly more than the individual $V_4$ scores of 3.13 and 3.22.)
Thus, we believe that this provides modest further support for the
hypothesis that increased $V_4$ scores corresponding to additional authors.

\subsection{Essays of known authorship}

Yet another way to test whether the $V_4$ statistic increases with multiple
authorship is to take documents of known authorship outside the judicial
realm (where clerks are not a factor) and combine them in different ways.

We considered the following historical essays:
\textit{Walden} by H. D. Thoreau (hereafter ``Walden'');
\textit{The Communist Manifesto} by K. Marx and F. Engels, in English
translation (``Manifest'');
\textit{On the Origin of Species} by C. Darwin (``Species'');
and \textit{On Liberty} by J. S. Mill (``Liberty'').
We divided Walden and Liberty into discrete 2000-word chunks (discarding
any leftover words), divided Manifest into discrete 1000-word chunks (since
it is shorter), and left Species as the 15 separate chapters in which it
was written.  We then tried combining them in different ways.  Our results
are in Table \ref{tab6}.

\begin{table}[b]
\tablewidth=245pt
\caption{Variability statistics when combining different essays together}\label{tab6}
\begin{tabular*}{245pt}{@{\extracolsep{\fill}}lcc@{}}
\hline
&  \textbf{\# units\phantom{0}} & $\bolds{V_4}$\\
\hline
Liberty & 23 & 1.799 \\
Manifest & 17 & 1.814 \\
Walden & 57 & 2.255 \\
Species & 16 & 2.999 \\ [3pt]
Liberty $+$ Manifest & 40 & 2.412 \\
Liberty $+$ Walden & 80 & 2.595 \\
Liberty $+$ Species & 39 & 3.830 \\
Manifest $+$ Walden & 74 & 2.548 \\
Manifest $+$ Species & 33 & 3.152 \\
Walden $+$ Species & 73 & 3.793 \\
Liberty $+$ Manifest $+$ Walden & 97 & 2.737 \\
Liberty $+$ Manifest $+$ Walden $+$ Species & 113\phantom{0} & 3.580 \\
\hline
\end{tabular*}
\end{table}

Once again, we see clear evidence that combining multiple authors leads
to larger $V_4$ scores, consistent with the hypothesis that larger $V_4$
scores indicate additional authors.  Indeed, in every case, the combined
$V_4$ is larger than any of the individual $V_4$ scores.

Again, this finding is fairly robust to subsampling.  For example,
considering just units \#1--9 of each collection (denoted by ``9''), we
obtain in Table~\ref{tab7}.\looseness=-1

\begin{table}
\tablewidth=220pt
\caption{Variability statistics when combining different essays together and
subsampling}\label{tab7}
\begin{tabular*}{220pt}{@{\extracolsep{\fill}}lcc@{}}
\hline
&  \textbf{\# units} & $\bolds{V_4}$ \\
\hline
Liberty9 & \phantom{0}9 & 1.717 \\
Manifest9 & \phantom{0}9 & 1.735 \\
Walden9 & \phantom{0}9 & 1.760 \\
Species9 & \phantom{0}9 & 3.240 \\ [3pt]
Liberty9 $+$ Manifest9 & 18 & 2.388 \\
Liberty9 $+$ Walden9 & 18 & 2.133 \\
Liberty9 $+$ Species9 & 18 & 3.873 \\
Manifest9 $+$ Walden9 & 18 & 2.568 \\
Manifest9 $+$ Species9 & 18 & 3.262 \\
Walden9 $+$ Species9 & 18 & 3.847 \\
\hline
\end{tabular*}
\end{table}

Even with this subsampling, the combined collections always have
larger $V_4$ scores than the individual collections, usually substantially
so.  Also, the results for the subsampled collections are generally
quite similar to the corresponding results for the full collections
(though Walden9 is rather less variable than Walden for some reason),
thus confirming that $V_4$ is largely unaffected by the size of a collection
but rather concentrates on the writing variability itself.

\subsection{Fictional novels}\label{sec-fiction}

For completeness, we also consider some famous historical
fictional novels (\href{http://www.gutenberg.org}{www.gutenberg.org}), namely:
\textit{Oliver Twist}, by C. Dickens (hereafter ``Oliver''),
\textit{The Three Musketeers} by A. Dumas (``Three''),
\textit{Pride and Prejudice} by J. Austen (``Pride''),
\textit{A Study in Scarlet} by A. C. Doyle (``Scarlet''),
and \textit{Alice in Wonderland} by L. Caroll (``Alice'').
Each novel was chopped into 2000-word units (again discarding any leftover).
The results are in Table \ref{tab8}.

\begin{table}
\tablewidth=210pt
\caption{Variability statistics when combining different fictional novels
together}\label{tab8}
\begin{tabular*}{210pt}{@{\extracolsep{\fill}}lcc@{}}
\hline
&  \textbf{\# units} & $\bolds{V_4}$\\
\hline
Pride & \phantom{0}60 & 1.730 \\
Alice & \phantom{0}13 & 1.815 \\
Oliver & \phantom{0}78 & 1.847 \\
Scarlet & \phantom{0}21 & 2.041 \\
Three & 114 & 2.058 \\ [3pt]
Pride $+$ Alice & \phantom{0}73 & 2.233 \\
Pride $+$ Oliver & 138 & 2.326 \\
Pride $+$ Scarlet & \phantom{0}81 & 2.160 \\
Pride $+$ Three & 174 & 2.310 \\
Alice $+$ Oliver & \phantom{0}91 & 2.025 \\
Alice $+$ Scarlet & \phantom{0}34 & 2.388 \\
Alice $+$ Three & 127 & 2.306 \\
Oliver $+$ Scarlet & \phantom{0}99 & 1.949 \\
Oliver $+$ Three & 192 & 2.191 \\
Scarlet $+$ Three & 135 & 2.179 \\
\hline
\end{tabular*}
\end{table}

Once again, the $V_4$ scores for the combined collections are larger than
the individual $V_4$ scores (with one exception: Oliver $+$ Scarlet),
sometimes substantially so.  Of course, it could be argued that
fictional writing is more free-form and thus has different stylometric
properties from such serious and formal writing as USSC judgments.
Nevertheless, these results do provide some sort of further support to
the hypothesis that larger $V_4$ scores indicate additional authors.

In the interests of fair reporting, we note that we also experimented
briefly with the novel \textit{War and Peace} by L. Tolstoy, broken up
into 281 different 2000-word chunks.  We found that this collection
had a surprisingly high $V_4$ score, 2.675, which did not significantly
increase (in fact, it sometimes even decreased) when combined with
other collections.  So, these results went against the hypothesis that
additional authors always leads to larger $V_4$ scores, perhaps due to the
unusually high variability of this novel itself.

Despite this caveat, overall we feel that the results of this section
provide modest additional support for the use of the $V_4$ statistic when
considering issues of multiple authorship.

\section{Authorship identification}

A related question is whether it is possible to identify which justice
is the (recorded) author of a judgment, based only on the writing style.
We posed this question to a small number of USSC constitutional
scholars.  The consensus was that while they could perhaps identify
authorship based on the case name or known passages, they could not do
so by writing style alone.  We now consider
the extent to which this identification
can be done by appropriate computer algorithms.
This question is thus similar in spirit to the Shakespeare authorship
question [Seletsky, Huang and Henderson-Frost (\citeyear{seletsky}); Burns~(\citeyear{burns}); Wikipedia (\citeyear{wikipedia})],
and also to the Federalist
Papers authorship question [Mosteller and Wallace (\citeyear{mosteller})]
and the Reagan radio address
analysis [Airoldi et al. (\citeyear{fienberg1}); Airoldi, Fienberg and Skinner (\citeyear{fienberg2})].  Of course, there is one
important difference here: in most instances the
recorded authorship of USSC judgments is
\textit{known}.  However, we still view this as a useful
test of the extent to which different USSC justices have identifiably
distinct writing styles.

We shall consider both naive Bayes classifiers and linear
classifiers, and shall see that each performs quite well at this task,
achieving success rates as high as 90\%.
(Other possible approaches include neural networks,
support vector machines, etc., but for simplicity we do not consider
them here.)

In each case, we shall consider a particular pair of justices (say,
Justices $A$ and $B$).  We shall consider the collection of all
USSC judgments whose recorded author is either $A$ or $B$, and shall
partition this collection into a disjoint training set and testing set.
Using only the training set, we develop a model for classifying judgments
as being authored by $A$ or $B$.  We then test to see if our model
classifies authorship correctly on the testing set.

\subsection{Naive Bayes classifier}

We begin with a naive Bayes classifier.  More specifically, we assume
that conditioned on the recorded author being Justice $A$, the conditional
distribution of the fraction $f_j$ of function word $j$ appearing in the
judgment is normally distributed.  (Of course, the normal distribution is
not the only choice here, and the ``true'' distribution is presumably a
rather complicated mixture, over the total number of words, of multinomial
distributions normalized by the total number of words in each judgment.
But the normal distribution appears to be a good enough approximation
for our purposes.)  We further assume that the corresponding mean and
variance are given by the sample mean and variance of all judgments by
Justice $A$ in the training set.  In addition, we assume (since we are
being ``naive'') that these different fractions $f_j$ (over different
function words $j$) are all conditionally independent.

Together with the uniform prior distribution
on whether the author is Justice $A$ or
$B$, this gives the log-likelihood for a given judgment being authored
by Justice~$A$, namely,
%
\begin{equation}\label{nbll}
\mathit{loglike}(A)  =  C - \sum_{j=1}^{\numwords} \biggl( \half \log(v_j)
+ (f_j-m_j)^2/2v_j \biggr)
\end{equation}
for some constant $C$,
where $f_j$ is the fraction
of words which are reference word~$j$ in the test judgment under
consideration, and
where $m_j$ and $v_j$ are the sample mean and variance of the fraction
of words which are reference word $j$, over all judgments in
the training set authored by $A$.

Similarly, we can compute $\mathit{loglike}(B)$.  The model then classifies
the test judgment as being authored by $A$ if $\mathit{loglike}(A) >
\mathit{loglike}(B)$, otherwise it classifies it as being authored by $B$.

\subsection{Linear classifier}

Another approach is a \textit{linear classifier}.  Specifically, let~$\T$
be a training set consisting of various judgments by $A$ or $B$,
with $|\T|=n$.  We consider the linear regression model
\[
\bY  =  \bx \beta + \epsilon,
\]
where $\epsilon$ is an $n \times 1$ vector of independent zero-mean errors.
Here $\bY$ is an $n \times 1$ vector of $\pm 1$, which equals
$-1$ for each judgment in the training set authored by~$A$, or
$+1$ for each judgment in the training set authored by $B$.
Also, $\bx$ is the $n \times 64$ matrix given by
\[
\bx  =  \pmatrix{
1 & f_{1,1} & f_{1,2} & \ldots & f_{1,\numwords} \cr
1 & f_{2,1} & f_{2,2} & \ldots & f_{2,\numwords} \cr
\vdots & \vdots & \vdots & & \vdots \cr
1 & f_{n,1} & f_{n,2} & \ldots & f_{n,\numwords}},
\]
where $f_{i,j}$ is the fraction of words in judgment $i$ (in the
training set) which are function word $j$.  For this model, the usual
least-squares estimate for $\beta$ (which corresponds to the MLE if
the $\epsilon_i$ are assumed to be iid normal) is given by
\[
\betah  =  ( \bx^T \bx )^{-1} \bx^T \bY.
\]

Once we have this estimate $\betah = (\betah_0,\betah_1,\ldots,\betah_n)$,
then given a fresh test
judgment having function word fractions $g_1,g_2,\ldots,g_{\numwords}$,
we can compute the linear fit value
\[
\ell  =  \betah_0 + \sum_{j=1}^{\numwords} \betah_j  g_j.
\]
Then, if $\ell<0$, we classify the test judgment as being authored
by $A$, otherwise we classify it as being authored by $B$.

Below we shall consider both the linear classifier and the naive Bayes
classifier.  We shall see that, generally speaking, the linear classifier
outperforms the naive Bayes classifier, sometimes significantly so.

\subsection{Testing accuracy via cross-validation}

To test the accuracy of our model, we use \textit{leave-one-out
cross-validation}.  That is, for each judgment by either~$A$~or~$B$,
we consider that one judgment be the test set, with all other
judgments by either $A$ or $B$ comprising the training set.  We then
see whether or not our model classifies the test judgment correctly.
Finally, we count the number of correct classifications, separately over
all judgments by $A$, and over all judgments by $B$.

\subsubsection{Results: Naive Bayes classifier}

We ran software [Rosenthal (\citeyear{software})] to perform the cross-validation test
using the naive Bayes classifier, for various pairs of justices $A$
and $B$.  Our results are in Table \ref{tab9}.

\begin{table}
\tablewidth=262pt
\caption{Authorship identification results using the naive Bayes classifier}\label{tab9}
\tabcolsep=3pt
\begin{tabular*}{262pt}{@{\extracolsep{\fill}}lccc@{}}
\hline
\textbf{Justice} $\bolds{A}$ & \textbf{Justice} $\bolds{B}$
& \hspace*{1pt}$\bolds{\operatorname{success}(A)}$ &\multicolumn{1}{c@{}}{\hspace*{1pt}$\bolds{\operatorname{success}(B)}$} \\
\hline
Scalia & Kennedy & $133/156 = 0.853$ & $129/147 = 0.878$ \\
Scalia & Souter & $132/156 = 0.846$ & $119/143 = 0.832$ \\
Scalia & Stevens & $130/156 = 0.833$ & $120/148 = 0.811$ \\
Scalia & Rehnquist & $139/156 = 0.891$ & $101/127 = 0.795$ \\
Kennedy & Souter & $139/147 = 0.946$ & $121/143 = 0.846$ \\
Kennedy & Stevens & $122/147 = 0.830$ & $113/148 = 0.764$ \\
Kennedy & Rehnquist & $124/147 = 0.844$ & \phantom{0}$97/127 = 0.764$ \\
Souter & Stevens & $118/143 = 0.825$ & $124/148 = 0.838$ \\
Rehnquist & Breyer & $111/127 = 0.874$ & $105/121 = 0.868$ \\
Rehnquist & Stevens & \phantom{0}$76/127 = 0.598$ & $113/148 = 0.764$ \\
Rehnquist & Thomas & \phantom{0}$76/127 = 0.598$ & \phantom{0}$94/140 = 0.671$ \\ [3pt]
Scalia & Scalia dissent & $140/156 = 0.897$ & \phantom{0}$72/108 = 0.667$ \\
Stevens & Stevens dissent & $122/148 = 0.824$ & $124/205 = 0.605$ \\
Scalia & Stevens dissent & $141/156 = 0.904$ & $118/205 = 0.576$ \\
\hline
\end{tabular*}
\end{table}

We see from these results that our naive Bayes classifier performs
fairly well on majority opinions,
often achieving a success rate over 80\%.  (This is
fairly consistent across all pairings, not just those shown
in Table \ref{tab9}; in particular, the success rate for majority
opinions is over 70\% for
all $\frac{8 \times 7}{2} = 28$
possible pairings except for five: Scalia--Thomas,
Souter--Stevens, Rehnquist--Stevens, Stevens--Thomas
and Rehnquist--Thomas.)  This appears to be quite a good performance,
especially considering the minimal assumptions that have gone into
the model.  (Presumably a more sophisticated model could achieve even
higher success rate.)  So, we see this as evidence that USSC judgment
authors can indeed be distinguished by their writing style, in fact just
by the pattern of fractions of function words used.

We also note that there is some variability concerning which justices'
writing styles are most easily distinguished.  For example, Rehnquist and
Breyer are apparently relatively easy to distinguish from one another,
while Rehnquist and Thomas are rather more difficult.

The algorithm does not perform as well on the dissenting opinions,
presumably because they tend to be shorter and thus less clearly
representative of their author's writing style.  In fact,
when comparing dissent to majority opinions, the algorithm tends to
classify too many judgments as being from the majority collection, and
this weakness remains whether the majority and minority collections are
from the same justice or from two different justices.

\begin{figure}

\includegraphics{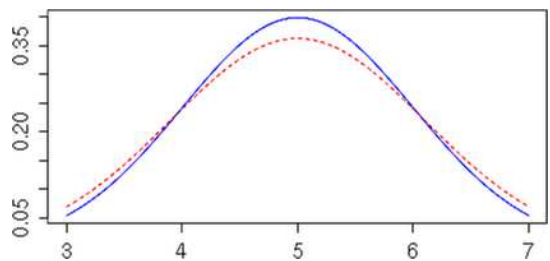}

\caption{Simple illustrative likelihood functions for
hypothetical justices A (solid line) and B (dashed line), for which
about 70\% of judgments would be classified as A's regardless of which
distribution they came from.}\label{fig3}
\end{figure}

\begin{remark*}
Our results above show some asymmetries, for example, there is much greater
success distinguishing Stevens' opinions from Rehnquist's (0.764) than
vice-versa (0.598).  This may seem counterintuitive but it provides no
contradiction.  For a simple illustration, if there were just one
function word, and $A$'s function word
distribution had mean 5 and standard deviation 1, while
$B$'s function word
distribution had mean 5 and standard deviation 1.1, then $A$'s likelihood
function would usually be above $B$'s (Figure \ref{fig3}), and about 70\% of
opinions would be classified as $A$'s regardless of which distribution
they came from.
\end{remark*}

\subsubsection{Results: Linear classifier}

We also ran software [Rosenthal (\citeyear{software})] to perform the cross-validation test
using a linear classifier, again for various pairs of justices $A$
and $B$.  Our results are in Table \ref{tab10}.

\begin{table}[b]\tablewidth=287pt
\caption{Authorship identification results using the linear classifier}\label{tab10}
\begin{tabular*}{287pt}{@{\extracolsep{\fill}}lccc@{}}
\hline
\textbf{Justice} $\bolds{A}$ & \textbf{Justice} $\bolds{B}$
&\hspace*{2pt}$\bolds{\operatorname{success}(A)}$ &\multicolumn{1}{c@{}}{\hspace*{2pt}$\bolds{\operatorname{success}(B)}$} \\
\hline
Scalia & Kennedy & $135/156 = 0.865$ & $135/147 = 0.918$ \\
Scalia & Souter & $137/156 = 0.878$ & $123/143 = 0.860$ \\
Scalia & Stevens & $125/156 = 0.801$ & $126/148 = 0.851$ \\
Scalia & Rehnquist & $137/156 = 0.878$ & $108/127 = 0.850$ \\
Kennedy & Souter & $138/147 = 0.939$ & $132/143 = 0.923$ \\
Kennedy & Stevens & $135/147 = 0.918$ & $128/148 = 0.865$ \\
Kennedy & Rehnquist & $133/147 = 0.905$ & $110/127 = 0.866$ \\
Souter & Stevens & $122/143 = 0.853$ & $131/148 = 0.885$ \\
Rehnquist & Breyer & $121/127 = 0.953$ & $110/121 = 0.909$ \\
Rehnquist & Stevens & \phantom{0}$88/127 = 0.693$ & $118/148 = 0.797$ \\
Rehnquist & Thomas & \phantom{0}$77/127 = 0.606$ & \phantom{0}$92/140 = 0.657$ \\ [3pt]
Scalia & Scalia dissent & $131/156 = 0.840$ & \phantom{0}$77/108 = 0.713$ \\
Stevens & Stevens dissent & \phantom{0}$99/148 = 0.669$ & $151/205 = 0.737$ \\
Scalia & Stevens dissent & $125/156 = 0.801$ & $164/205 = 0.800$ \\
\hline
\end{tabular*}
\end{table}

Comparing these results with those from the previous subsection shows
that the linear classifier performs even better than the naive Bayes
classifier, with success rates often close to 90\%.  (This is again
fairly consistent across all pairings; in particular, the success rate is
above 80\% for all possible majority opinion pairings with the exception
of Rehnquist--Stevens and those involving Thomas.)  This provides further,
even stronger evidence that it is indeed possible to distinguish between
different USSC justices' judgments solely on the basis of writing style.

Once again, there is some variability concerning which justices' writing
styles are most easily distinguished.  For example, success rates for
distinguishing Rehnquist from Breyer are over 90\%, while those for
distinguishing Rehnquist and Thomas are in the 60s.

On dissenting opinions, the linear classifier appears to be
less prone to incorrectly classifying almost all judgments as being from
the majority opinion collection.  Rather, it is better balanced between
the two collections.  However, it still finds the dissenting opinions
to be challenging, with success rates ranging from 84\% (quite good)
down to 67\% (rather poor).

\begin{remark*}
It is possible to examine the regression coefficients to see which words
are most used to distinguish justices.  For example, when comparing
Kennedy to Scalia, the regression coefficient for the function word
\textit{now} is $-540$, while that for \textit{such} is $+204$.  And, indeed,
Kennedy's judgments use \textit{now} over twice as frequently as Scalia's,
but use \textit{such} less than half as frequently.
\end{remark*}

\subsection{Outlier detection}

Finally, we briefly note that the above naive Bayes approach can easily
be adapted to the issue of \textit{outlier detection}.  Suppose a collection
of $n$ judgments is given, and it is believed that they were all written
by the same author with one exception (e.g., perhaps a justice allowed
his clerks to write just one of his opinions each term, something we
may explore more fully in separate work).  That is, there are $n-1$
``decoy'' judgments all written by the same author, plus one unknown
``test'' judgment having different authorship.  In this case, for each
individual judgment, we proceed by excluding that judgment, computing
sample means $m_j$ and variances $v_j$ for each reference word $j$ based
on the other $n-1$ judgments, and then computing a log-likelihood for the
individual judgment as in~(\ref{nbll}).  The higher this log-likelihood
value, the better the individual judgment ``fits in'' with all the other
judgments.  We can then rank all the individual judgments from 1 to $n$
in terms of their log-likelihood scores, from smallest log-likelihood
(i.e., most likely to be the outlier) to largest log-likelihood (i.e.,
least likely to be the outlier).

To score the performance of such outlier detection, suppose our
algorithm gives the true outlier a rank of $i$.  If $i=1$, the algorithm
has performed perfectly, while if $i=n$, then the algorithm has
completely failed.  So, we can convert this to a score from 0 (worst) to
100 (best), by the simple linear transformation
%
\begin{equation}\label{outlierscore}
\mathit{score}  =  \frac{n-i}{n-1} \times 100.
\end{equation}
To test this algorithm, we averaged the $\mathit{score}$ (\ref{outlierscore}) over a
collection of test judgments, to compute a final average score between 0
(worst) and 100 (best).  We used the following collections: Scalia's
156 judgments considered herein;
Kennedy's 147 judgments considered herein;
Rehnquist's 127 judgments considered herein;
the 24 judgments from Volume 8 (1807--1808) of the USSC (obtained
by extending our software [Rosenthal (\citeyear{software})] to download older USSC judgments
from the Justia~(\citeyear{justia}) web site); and the 114 segments of \textit{The
Three Musketeers} (``Three'') as discussed in Section \ref{sec-fiction}.
Our results are in Table \ref{tab11}.

\begin{table}[b]
\tablewidth=230pt
\caption{Outlier detection results}\label{tab11}
\begin{tabular*}{230pt}{@{\extracolsep{\fill}}lcc@{}}
\hline
\textbf{Test collection} & \textbf{Decoy collection} & \textbf{Average score}
\\
\hline
Scalia & Three & 99.44 \\
Three & Scalia & 99.87 \\
Scalia & Volume 8 & 58.55 \\
Volume 8 & Scalia & 99.71 \\
Scalia & Kennedy & 64.41\\
Kennedy & Scalia & 65.72 \\
Scalia & Rehnquist & 57.75 \\
Rehnquist & Scalia & 71.11 \\
\hline
\end{tabular*}
\end{table}

We see that the algorithm can very easily distinguish the fictional
work \textit{The Three Musketeers} from such serious writings as Scalia's
USSC judgments.  Furthermore, it can easily pick out an old Volume 8
judgments from a sea of modern Scalia judgments.  Interestingly, this
last result is highly asymmetric (even more so than that suggested by
Figure \ref{fig3}), that is, the algorithm is much worse
at picking out a single Scalia judgment from a sea of Volume 8 judgments.
Perhaps unsurprisingly, the algorithm has less success picking out a
single Scalia judgment from a sea of Kennedy judgments, or vice versa.
Indeed, its scores, near 65, are only moderately better than pure chance
guessing (which would produce an average score of 50).  For Scalia
versus Rehnquist---two justices who apparently write their own opinions---the
scores are not much better.  This illustrates that it is easier to
identify judgment authorship when given two large collections, than when
given a single large collection with just one outlier.

\section{Summary}

In this paper we have presented methodology and software for
investigations of USSC judgments by using statistical properties of
function words.

First, we have investigated the variability of writing style over
various collections of judgments, in particular, of majority decisions
written by different justices.  We have seen that it is possible to
uncover statistically significant evidence that one USSC justice (e.g.,
Kennedy) has greater writing-style variability than another justice
(e.g., Scalia), which may indicate that the first justice relies on law
clerk assistance to a greater extent than does the second justice.

Second, we have investigated the extent to which unknown authorship of
USSC judgments can be determined based solely on function word statistics.
We have seen that both naive Bayes classifiers and linear classifiers
perform fairly well at this task, achieving cross-validation success
rates approaching 90\%.  While authorship is typically known for all USSC
opinions, our approach reveals that justices---even with contributions by
clerks---have writing styles which are distinguishable from one another.
(In a different direction, one could perhaps use function words to
identify authorship for the handful of \textit{per curiam} decisions in which
the Court does not reveal authorship, though we do not pursue that here.)

Of course, our approach---or any textual analysis---can provide only
circumstantial evidence of collaborative authorship, not definitive proof.
A low $V_4$ score can reflect that a justice does her own writing, that the
justice closely edits her clerks' work, or that the clerks are all highly
effective at mimicking their justices' writing style; these states of the
world are observationally equivalent.  However, we do believe that our
results provide compelling evidence that justices over time are indeed
relying more on their law clerks, and that justices vary considerably
from one another in this regard.

Overall, we hope that the methodology and software [Rosenthal (\citeyear{software})]
presented here will provide useful insights into USSC writings, as well
as a helpful starting point for other statistical investigations into
other bodies of writing in other contexts.

\section*{Acknowledgments}
We thank the editors and referees for very helpful comments that
greatly improved our manuscript.

\printaddresses

\end{document}